\documentclass[aps,pre,showpacs,twocolumn,superscriptaddress]{revtex4}
\usepackage{epsfig}
\usepackage{graphicx}
\usepackage{amsmath}
\usepackage{amssymb}
\usepackage{verbatim}   
\usepackage{color}      
\usepackage{subfigure}  

\newcommand{\beq}{\begin{equation}}
\newcommand{\eeq}{\end{equation}}

\begin{document}

\title{Link and subgraph likelihoods in random undirected networks with fixed and partially
  fixed degree sequence} \date{\today} \author{Jacob G. Foster}
\affiliation{Complexity Science Group, Department of Physics and
  Astronomy, University of Calgary, Calgary, Canada} \author{David V.
  Foster} \affiliation{Institute for Biocomplexity and Bioinformatics,
  University of Calgary, Calgary, Canada } \author{Peter Grassberger}
\affiliation{Complexity Science Group, Department of Physics and
  Astronomy, University of Calgary, Calgary, Canada}
\affiliation{Institute for Biocomplexity and Bioinformatics,
  University of Calgary, Calgary, Canada } 
  \author{Maya Paczuski} \affiliation{Complexity Science
  Group, Department of Physics and Astronomy, University of Calgary,
  Calgary, Canada}

\begin{minipage}[t]{7.0in}
\begin{abstract}

  The simplest null models for networks, used to distinguish significant features
  of a particular network from {\it a priori} expected features, are random
  ensembles with the degree sequence fixed by the specific network of interest. These
  ``fixed degree sequence" (FDS) ensembles are, however, famously resistant to analytic 
  attack. In this paper we introduce ensembles with partially-fixed degree sequences
  (PFDS) and compare analytic results obtained for them with Monte Carlo results for 
  the FDS ensemble. These results include link likelihoods, subgraph likelihoods, and
  degree correlations. We find that local structural features in the FDS ensemble can
  be reasonably well estimated by simultaneously fixing only the degrees of few nodes,
  in addition to the total number of nodes and links. As test cases we use a food 
  web, two protein interaction networks  (\textit{E. coli, S. cerevisiae}), the 
  internet on the autonomous system (AS) level, and the World Wide Web. Fixing just the 
  degrees of two nodes gives the mean neighbor degree as a function of node degree, 
  $\langle k'\rangle_k$, in agreement with results explicitly obtained from rewiring.
  For power law degree distributions, we derive the disassortativity analytically. 
  In the PFDS ensemble the partition function can be expanded diagrammatically.
  We obtain an explicit expression for the link likelihood to lowest order, which 
  reduces in the limit of large, sparse undirected networks with $L$ links and with
  $k_{\rm max} \ll L$ to the 
  simple formula $P(k,k') = kk'/(2L + kk')$.  In a similar limit, the probability for 
  three nodes to be linked into a triangle reduces to the factorized expression
  $P_{\Delta}(k_1,k_2,k_3) = P(k_1,k_2)P(k_1,k_3)P(k_2,k_3)$.

\end{abstract}

\pacs{02.50.Cw, 02.70.Uu, 05.20.Gg, 87.10.+e, 87.23.Cc, 89.75.Fb, 89.75.Hc}
\end{minipage}

\maketitle

\section{Introduction}

A pivotal question of empiricism is the degree to which
the results of an observation are expected. In ideal cases, either
predictions based on these expectations remain valid in view of new 
measurements, or the expectations have 
to be changed. But this clear distinction is often blurred by 
uncertainties resulting from measurement errors, imprecision of
model parameters, or the impossibility of extracting exact predictions 
from complicated models.  Whether or not the problem at 
hand is  a typical instance of a wider class of problems that
are already understood is a question of statistical 
inference. In rare cases, the consequences of the expectations (or the 
model) can be derived analytically prior to observation. If 
this is not feasible, a widely used strategy is to 
construct a large number of ``surrogates''~\cite{eubank}, or
instances of a well-defined null model encapsulating the 
expectations, and to compare the actual observations to this artificial data.

Constructing surrogates is equivalent to simulating a statistical
ensemble. In choosing weights for the ensemble of surrogates one often uses 
Occam's razor---no outcome compatible with the null hypothesis should be
preferred, and all such outcomes are equally likely.  This is similar
to Jaynes' construction of statistical mechanics by maximizing Shannon
entropy with physically meaningful constraints. Consequently, the
numerical construction of surrogates often uses Monte Carlo
methods~\cite{schmitz-schreiber} similar to those used in statistical
mechanics.

This paper addresses properties of ensembles used as null
models for complex networks. Predictions based on the null models
fix expectations, and thereby determine whether or not  deviations
in the properties of an actual network 
are functionally or historically significant. While the 
numerical construction of surrogates of these ensembles has received  
attention in the recent literature~\cite{maslov-sneppen,milo04,chen05}, much 
less is known about analytic methods (see discussion below).

Nowadays networks attract enormous interest as representations of
complex systems. They take various guises
in biological, social, technological and physical contexts. The nodes
designate distinct degrees of freedom (e.g agents, species,
genes, magnetic concentrations in the solar photosphere, or earthquakes) 
and the links identify primary interactions or relationships between 
pairs of nodes (e.g. co-authorship, predator-prey relations, gene regulation, 
magnetic flux tubes, or seismic correlations). For examples see~\cite{newman03,
newmancollab,aids,gonorrhea,foodweb,stu,mayapeterquake,abc}. The ubiquity 
of networks and their relatively easy visualization as graphs, together 
with notions of universality prevalent in the physics community, have driven
speculations that the structure of networks can shed light
on fundamental principles of social or biological organization, such
as political behavior, ecosystem dynamics, brain function
or the regulated homeostasis of organisms.

At the simplest level, networks are purely static entities, with each
pair of distinct nodes connected by no more than one edge (or ``link"). 
If, in addition, the interaction {\it strength} is disregarded (which
often is a very useful simplification) the adjacency matrix $\cal M$
for the graph is a square (0,1) matrix. If $M_{ij}=1$ then an edge
points from node $i$ to node $j$; if $M_{ij}=0$ then the edge is
absent.  Without self-interactions, $M_{ii}=0$. For undirected networks, 
the adjacency matrix is symmetric, $M_{ij}=M_{ji}$. The {\it degree}
$k_i$ of node $i$ is then defined as the number of edges incident on it, 
$k_i=\sum_jM_{ji}$. Several reviews may be found in 
Refs.~\cite{albert02,dorogo02,newman03}.

Section II defines more precisely the network ensembles (or null models) 
we consider in this paper. Our analytical methods focus on ensembles 
where the total number of links and nodes in the network is specified
as well as the degrees of a small subset of nodes. These are called 
ensembles of ``partially fixed degree sequence'' (PFDS). Analytic 
predictions based on the PFDS ensembles can be compared with numerical 
results from a `rewiring' algorithm for ensembles with fixed degree 
sequence (FDS), where the number of links attached to \emph{every} node 
in the network is simultaneously specified.  Section III mainly
recalls previous results. We review Monte Carlo methods for
sampling the FDS ensemble.  Then we discuss how such null models can
be used, and we conclude by recalling previous analytic approaches.
Section IV discusses some results derived later in Section~V, namely 
analytic estimates of the link linkelihood $p_{ij}$ (the linkelihood for 
a link to connect nodes $i$ and $j$). It uses them to make predictions 
for the average nearest neighbor degree $\langle k'\rangle_k$ and for 
disassortativity. The calculation gives an excellent description 
of $\langle k'\rangle_k$ for large $k$,
e.g. for an \textit{Escherichia coli} protein interaction network and an AS
level map  of the Internet. We also compute $\langle k'\rangle_k$ analytically
for the case where the degree distribution is a power law, using
Eq.~(\ref{Pkk}) given below. In that case, the naive approximation for 
$P(k,k')$ would give divergent or ill-defined results.

Section V contains our main analytic results. In order to keep the notational 
clutter of this section to a minimum and to emphasize the intuitive nature of 
the results, most intermediate steps are moved to Appendix B. 
In the limit of large sparse networks with $L$ links and with the maximal
degree much less than $L$, we find that the link
likelihood depends only on $L$ and on the degrees $k$ and $k'$ of the two nodes,
and is given by
\beq
    P(k,k') \approx kk'/(2L + kk')\; .            \label{Pkk}
\eeq
This improves substantially over the widely used `naive' approximation $P(k,k') 
\approx kk'/2L$. We also find that the disassortativity 
of the FDS ensembles corresponding to several real world networks is 
well-described by PFDS ensembles simultaneously fixing the degrees of two nodes
at a time.  Finally, we find an expression for 
the likelihood of a triangle, which factorizes in the same limit of large 
sparse networks (and when all three degrees are much larger than 1) to 
$P_{\Delta}(k_1,k_2,k_3) = P(k_1,k_2)P(k_1,k_3)P(k_2,k_3)$, with $P(k,k')$ given 
again by Eq.~(\ref{Pkk}). The paper ends in section VI with a discussion and 
an outlook to further problems.

\section{Null Models for Networks}

\subsection{Erd{\"o}s-Renyi (Undirected) Graphs}

The simplest null hypothesis is that a given network is completely
random, not even the number of links being specified. The only
constraint is on the number of nodes, which is assumed to be $N$. Each
pair of nodes may be joined with at most one link.  Hence, the number
of labelled undirected graphs with fixed $N$ is $Z_0(N) =
2^{N(N-1)/2}$.  This quantity is the number of ways undirected links may be
placed in $\binom{N}{2}=N(N-1)/2$ possible positions. A statistical
ensemble is obtained by assigning weights to each graph. The most
natural choice is to weigh each graph with $L$ links by a factor
$p^L$, where $p$ is the probability that two given nodes are connected
by a link. This gives the average number of links as ${\bar L}= p
N(N-1)/2$. The average degree of a node, i.e. the average number of
links attached to it, is then ${\bar k}=2{\bar L}/N =p(N-1)$, and the
degree distribution is binomial. In the limit of sparse networks,
where $p\to 0$ for $N\to \infty$ such that $pN \to  const. $, the
degree distribution (or the probability $P(k)$ that a node has $k$
links) becomes Poissonian,

\beq
    P(k) =\frac {{\bar k}^k\exp^{-\bar k}}{k!} \;.       \label{Poisson}
\eeq

While this ensemble can be viewed as a ``grand canonical'' version of
the Erd{\"o}s-Renyi ensemble \cite{bollobas} since the particle fugacity
is fixed, it is more customary to
associate Erd{\"o}s-Renyi graphs with a different ensemble where the total
number $L$ of links is fixed, rather than just the average $\bar L$.
Park and Newman refer to the ensemble with fixed $L$ as
``canonical''~\cite{park03,park04},  making an analogy between the
number of links and the number of particles in traditional statistical
mechanics. However, we shall refer to this ensemble, and ensembles with similar
hard degree constraints, as microcanonical.

Excluding self-connections as well as multiple edges between any pair of
nodes gives
\begin{equation}
   Z_1(N,L) = \sum_{\left\{lg, N\right\}}^{\left\{L\right\}} 1 = \binom{\binom{N}{2}}{L} \qquad
\label{eq:simple}
\end{equation}
distinct labelled, undirected graphs with fixed $N$ and $L$
\cite{footnote1}.  The subscript ``$\left\{lg, N\right\}$'' indicates
a sum over labelled graphs with $N$ nodes, while the superscript
$\left\{L\right\}$ on the sum indicates, as in later formulae, the
constraints on the edges. The subscript ``1'' on $Z$ indicates that
$Z_{1}$ is the number of undirected graphs with one (global) hard
constraint on the links, just as $Z_0(N)$ is the number of undirected
graphs with zero constraints on the links.

With no further knowledge or constraints on the network, Occam's razor
suggests assigning equal weight to each labelled graph satisfying all
the hard constraints.  This corresponds exactly to the construction of 
microcanonical ensembles in statistical mechanics. For $Z_1$,
 each node has equal probability to be connected to any other
node. It is easy to show~\cite{burda05} that the distribution for the
number of links attached to each node is again Poissonian for sparse
networks with large $N$, where the grand canonical and microcanonical
ensembles become equivalent.

In contrast, observations of real networks reveal fat-tailed degree
distributions, which differ starkly from the situation where each
node has equal likelihood to be connected to any other node.  The most
salient consequence is that the average degree $\bar k$ fails to
characterize the connectivity of the nodes; in particular it cannot
account for the dominant nodes or ``hubs'' with many links, which
would not typically appear in the Erd{\"o}s-Renyi ensemble.

\subsection{Ensembles with Fixed Degree Sequences}

As a result, attention has moved to ensembles that build additional
information into the null hypothesis about the ``distinguishability''
or diversity of the nodes. Although many different and equally
plausible ways to account for diversity can be imagined, to begin we
focus on the most popular contemporary method.  This uses the random
ensemble of labelled graphs with fixed degree sequence (FDS) as the
relevant null model. The complete degree sequence simultaneously fixes
\emph{all} the one-node properties for each member of the ensemble, 
without reference to their
relationships (or links) in the network.  Obviously, it is
straightforward to obtain the degree sequence for any network, and
there exist numerical methods to estimate characteristic properties of
the corresponding FDS ensemble (see section III).

The microcanonical FDS ensemble is specified by assigning a specific
degree (= number of links) to each node, $k_i$ for $i= 1,..., N$, and
giving equal weight to each graph with this degree sequence, while
giving zero weight to all those graphs which have a different degree
sequence.  The
null hypothesis for any observable pertinent to a specific graph
${\cal G}$ with adjacency matrix ${\cal M}_{\cal G}$ is then obtained 
by taking its expectation value in the FDS ensemble with the same 
degree sequence.  For undirected graphs excluding self-interactions, 
the FDS partition sum is the number of symmetric (0,1) matrices with
zeroes on the diagonal and with fixed marginal sums, which can be
written according to our previous convention as \beq Z_N(N,L, k_1,
k_2, ..., k_{N-1}) = \sum_{\left\{lg, N \right\}}^{\left\{k_1...,
    k_N\right\}} 1 \;.  \eeq with $k_1 + k_2+ ...+ k_N=2L$. For most
networks of physical interest, $Z_N$ is astronomically large compared
to one, but vanishingly small compared to $Z_1$.  For instance, Chen
{\it et al.} \cite{chen05} numerically estimate the number of
$12\times 12$ (0,1) matrices with each row and column sum equal to 2
(and with no restrictions of symmetry or vanishing diagonal) to be
$\approx 2.196 \times 10^{16}$, which agrees well with the exact
number found by Wang and Zhang \cite{wang98}.  This number is much
smaller than the number of all $12\times 12$ $(0,1)$ matrices with 24
ones, which is $\binom{12^2}{24}\approx 1.69 \times 10^{29}$.  Despite
efforts by these and other mathematicians over decades
\cite{bender,wang98}, no well-developed, exact analytical approaches
are known for these combinatorial problems, but advanced computational
methods exist, as described in Section III.

\subsection{Partially Fixed Degree Sequences}

On the one hand, the difficulty of enumerating the number of graphs
in the FDS ensemble suggests strong correlations in
the graphs, since similar problems in systems lacking correlations can
often be solved exactly. Indeed, the FDS ensemble makes very
different predictions from the Erd{\"os}-Renyi (ER) ensemble, showing
that taking into account \emph{some} information about the nodes'
degrees is crucial.

On the other hand, it might be the case that not all the constraints
in the FDS ensemble must be taken into account simultaneously. After
all it is the simultaneous fixing of \emph{all} the marginal sums in
the matrix that makes the calculation of $Z_N$ difficult.  Perhaps
taking into account all the constraints, but not requiring them to be
simultaneously enforced, is already sufficient to capture some
nontrivial aspects of the FDS ensemble.  If this is possible, then we
must also identify which specific small subset of the nodes' degrees gives the most
reliable estimate of various expectation values in the FDS ensemble.

Here, we study ensembles where the degrees of a very small subset of
the nodes are simultaneously fixed -- the other degrees being arbitrary 
up to the constraint on the total number of links. We also demand that 
no more than one link may connect any two nodes in the network, and disallow
self-connections.  All graphs satisfying these constraints have equal
weight.  Those not satisfying these constraints are given zero weight.
These ensembles can all be viewed as sub-ensembles of the ER ensemble.
For each possible degree subset, we calculate the different partition 
functions corresponding to all possible subgraphs of a certain size. 
From these we can approximate
expectation values of various quantities in the FDS ensemble.

In the following we shall always label the nodes such that the first $m$
degrees, $k_1, ... k_m$, are fixed. We call the resulting ensembles
PFDS($m+1$): partially fixed degree sequence with $m+1$ constraints (the 
final constraint comes from fixing the number of links, $L$).
Clearly, putting more constraints on the ensemble of labelled graphs
diminishes its size until, when each and every edge is specified, the
ensemble contains just one member - the real network ${\cal G}$ being
studied.  For ensembles with increasing numbers of link constraints
this implies that $Z_{m+1}$ decreases monotonically with $m+1$, and
\begin{multline}
   1\equiv Z_{\cal G} \ll  Z_N(N,L, k_1, k_2, ..., k_{N-1})     \\
      \ll  Z_{m+1}(N,L, k_1, ... k_m) \ll Z_1(N,L)  \;\; , 
\end{multline}
for $1<m<N-1$. 

We find that fixing only the degrees of the nodes participating in the
small subgraph (e.g. link or triangle) under consideration, with
explicit exclusion of self-connection and multiple-connections between
any nodes, already gives a good approximation to the disassortativity
(and to other properties) in the FDS ensemble. As noted above, this uses 
information about the whole degree sequence, but in each contribution 
corresponding to one specific (labelled) subgraph only 
part of this information is used. 

The information stored in the degree sequence is most important when 
its distribution is very wide. Even for networks exhibiting broad degree distributions, 
such as protein interaction networks or autonomous system maps of the 
Internet, it is sufficient to fix the degrees of the node pairs directly 
involved (as well as the total number of links in the network) to obtain 
a good estimate of $\langle k'\rangle_k$ and of the disassortativity. In 
order to estimate the number of triangles (i.e. the clustering), one has 
to fix the degrees of node triples. If we fix in addition the degrees 
of (some) hubs, this slightly improves the approximations. It is much 
easier to make analytical calculations for small $m$, with the smallest
meaningful $m$ being the size of the subgraph being considered. Hence for link 
likelihoods and for $\langle k'\rangle_k$ this minimum is $m=2$, while
for triangle likelihoods it is $m=3$. By comparing
analytic properties of the PFDS($m$) ensemble with numerical estimates
of the FDS ensemble, we can assess to what extent the correlations in
the PFDS ensembles resemble those in the FDS and in the ER ensembles.

To begin, we will focus in Section IV on the link likelihood, $p_{ij}$, 
which is the probability that a randomly chosen graph from the ensemble 
contains an edge from node $i$ to node $j$. From this microscopic quantity 
one can calculate the standard degree-degree correlations that are commonly 
compared with real-world networks to identify statistically significant 
features. Details of the calculation of $p_{ij}$ are deferred to Section V 
(and to Appendix B). There we will also treat the generalization 
to $m\ge 3$ which is needed in order to estimate the frequencies of 
higher order quantities such as motifs~\cite{milo02}. As an example of 
a motif calculation, we include an estimate of the number of triangles.

\section{Background}

\subsection{Monte Carlo algorithms to estimate the FDS ensemble}
    \label{MC}

As for many other problems where one wants to sample complex
instances from some well-defined ensemble, here two basic
strategies predominate: Markov chain Monte Carlo and sequential
sampling \cite{Liu,milo04,PERM}.  For the present case, the most
obvious and popular Markov chain algorithm is the {\it rewiring
  algorithm} \cite{besag89,rao96,roberts00,cobb03,maslov-sneppen}.
We describe it here for directed graphs; the generalization to
undirected graphs is immediate.  We start by making an initial
network with $N$ nodes, no self-connections, and the desired
degree sequence, but without paying attention to multiple links
between pairs of nodes \cite{footnote2}.  The Monte Carlo
algorithm proper consists of a sequence of moves, randomly chosen
from a move set, which continues until equilibrium (i.e.
uniformity of sampling) is reached with sufficient accuracy. A
move is initiated by choosing randomly four different nodes
$i,j,l$ and $m$ with $M_{ij} >0$ and $M_{lm} >0$. If either 
$M_{im} >0$ or $M_{lj} >0$, a null move is performed (the graph 
is left as it is).  If neither of the pairs $\{i,m\}$ and $\{l,j\}$ were
already connected by a link, $M_{ij}$ and $M_{lm}$ are each decreased
by 1, while $M_{im}$ and $M_{lj}$ are increased by one. This
corresponds to swapping one pair of links. 

It can be shown easily that this algorithm eventually leads to a
graph without multiple links (provided such a graph consistent
with the fixed degree sequence exists). After this happens, the algorithm
satisfies detailed balance (any sequence of moves is equally likely to
be chosen as its reversed sequence) and is ergodic (each graph with
the same degree sequences can be reached by a suitable move sequence.)
As shown in~\cite{rao96}, ergodicity is not strictly satisfied, but
the few exceptions can be taken into account by including three-link
exchanges in the move set.

Sequential sampling proceeds, in contrast, by repeatedly building a
new graph from scratch. For this we start with an empty adjacency
matrix and fill its entries randomly. In the simplest version, this is
done without paying any attention to the degree sequence, to the
absence of self loops, or to the exclusion of multiple links. Instead,
the candidate graph is discarded if any of these constraints are
violated. In this way the uniformity of the sampling is guaranteed,
but the {\it attrition} (i.e. the chance to reach an illegal
configuration) is overwhelming, rendering the algorithm useless.  But
there are more sophisticated options for sequential sampling.
The most efficient algorithm studied in the literature~\cite{chen05}
uses detailed mathematical results for the structure of legal
adjacency matrices~\cite{Ryser57} to bias the matchings in a much more
clever way.

\subsection{Uses of null models}

Statistically significant deviations between a null model and a real
network point to organizing laws or historical accidents that are not
accounted for by the null hypothesis.  On the other hand, finding no
statistically significant deviations would promote the belief that the
entire structure of the network could be accounted for by the model,
e.g. by the complete degree sequence in case of the FDS ensemble.
This process of building null models can, in principle, be iterated to
understand the full set of organizing principles or physical
constraints on a network: one builds a null model, tests for
significant deviations, and then builds a new null model with richer
structure to try to reduce any significant deviations to typicality.

Through an application of this discriminatory method, Maslov {\it et
  al.}~\cite{maslov04} showed that a significant part of the
dissortativity~\cite{newman02} observed in the Internet could be
attributed to the broad degree distribution together with the
restriction of no multiple links between any pair of nodes.  For a
scale free network of $N$ nodes with a degree distribution $P(k) \sim
k^{-\gamma}$, the maximum expected degree $k_c(N)$ scales as $k_c(N)
\sim N^{1/(\gamma -1)}$. In a random network with no constraints on
edge multiplicities, the expected number of edges between the two
largest hubs would then scale as $k_c(N)^2/N \sim N^{2/(\gamma -1)
  -1}$. For $\gamma < 3$, this number diverges with $N$.  If the
constraint of no multiple edges is imposed, these links must be
distributed so that they connect the hubs to other nodes. This creates
fewer links between hubs than naively expected, and more links between
hubs and nodes with small degree; it also leads to a suppression of
links between nodes with small degrees (as the degrees of these nodes
are ``used up'' by connecting to hubs). The net effect is that fixing
a broad degree sequence decreases assortativity (the preference for
nodes with similar degrees to be connected to each
other)~\cite{maslov04}.

On the other hand, Milo {\it et al.}~\cite{milo02} have discovered 
subgraphs or motifs that are significantly more frequent in actual 
networks than in the corresponding FDS ensemble. Identifying 
these motifs allows a classification of networks that share the
same motifs. For instance feed-forward loops are overrepresented 
in gene regulation networks and in some electronic circuits, while fully 
connected triangles are most overrepresented in the world wide web.

So, on the one hand we see that the FDS ensemble, together with
non-trivial (power law) degree distributions, allows both
discrimination between features of the network and comparison with
other networks; on the other hand, the ensemble itself exhibits strong
correlations.  To explain how these correlations are related to each
other and to the degree sequence it is useful to have an analytic
approach. This is also important if one wants to develop more refined 
null models, or study very large networks for which rewiring is prohibitive. 
While most authors have considered the FDS ensemble as the most 
natural null model for networks, there have been attempts to
generalize to more complex ensembles. Maybe the most interesting is
due to Mahadevan {\it et al.}~\cite{mahadevan06}.

\subsection{Previous analytic approaches}

The present paper builds on a paper by Burda {\it et
  al.}~\cite{burda05}.  An alternative strategy to incorporate
information on degree distributions was proposed by Park and Newman
\cite{park03,park04}. While we fix the degree sequence $\{k(m),
m=0,1,\ldots\}$ {\it exactly}, Park and Newman constrain only the {\it
  average} numbers $\langle k(m)\rangle $, averaged over the ensemble.
Thus, while our approach is microcanonical, the one
of~\cite{park03,park04} is grand canonical. As in statistical
mechanics, calculations are often simpler in the grand canonical
ensemble, but they are feasible and not too difficult for the PFDS($m$)
ensemble considered in this paper, with $m$ small.  Note that for finite
sized networks, the two ensembles are not equivalent. Further, for a 
given network, physical arguments may suggest that one ensemble is more 
explanatory than another.

\section{Link likelihoods $p_{ij}$ and disassortativity in null models}

\begin{figure}
  \begin{center}
   \psfig{file=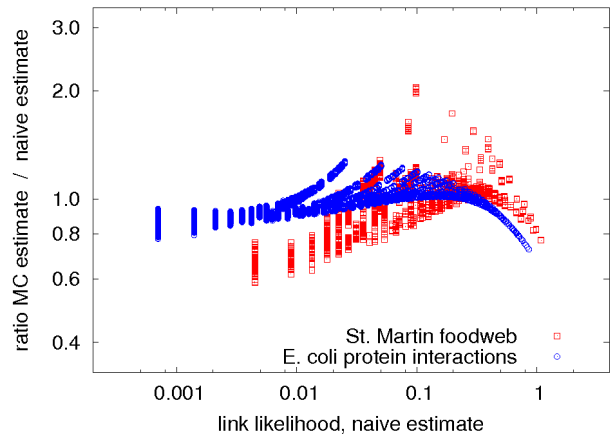,width=6.cm,angle=270}
   \caption{Log-log scatter plot of the naive analytic estimate of 
      link likelihood $k_{out,i}k_{in,j}/L$ (directed network, St. Martin 
      foodweb \cite{stmartin}) or $k_i k_j/2L$ (undirected network, 
      \textit{Eschericia coli} protein interaction network \cite{ecoli}) 
      versus the ratio of the Monte-Carlo rewiring estimate to the naive 
      estimate for the corresponding nodes. Note that the directed network 
      has considerably more scatter for given $k_{out,i},k_{in,j}$.} 
\label{scatter}
\end{center}
\end{figure}

For undirected networks, all pairs of nodes with the same degree 
have the same likelihood to be connected in the FDS ensemble. For
directed networks the likelihood to form a link from node $i$ with
out-degree $k_{out,i}$ to node $j$ with in-degree $k_{in,j}$ also
depends on $k_{in,i}$ and $k_{out,j}$. This is demonstrated in
Fig.~\ref{scatter}, where the actual link likelihood estimated using a
rewiring algorithm is plotted vs. the naive analytic estimate of the
link likelihood $k_{out,i}k_{in,j}/L$ (for pairs in the directed
networks) or $k_i k_j/2L$ (for pairs in the undirected network).  In
particular, directed networks exhibit a high degree of scatter for the
same values of the connected (out- and in-) degrees, showing the
importance of the other degrees associated with the pair (in- and
out-, respectively). Further, the likelihood does not approach the
naive estimate for $k_i k_j \ll L$. This is due to the constraint of
one link between two nodes and to the presence of hubs, which thus have 
to distribute their links to different nodes.

For any ensemble $A$, let $N_A(k,k')$ be the average number of links 
between nodes with degree $k$ and nodes with degree $k'$.  
In terms of the link likelihood,  
\beq
   N_A(k,k')= \sum_{i,j} p_{ij,A}\delta(k_i - k)\delta(k_j -k') \quad 
                       \label{NA}
\eeq
where the sum over $i,j$ indicates a sum over all pairs of nodes, and
$p_{ij,A}$ is the link likelihood for ensemble $A$. If the ensemble $A$
is the trivial ensemble consisting of just one network, namely the  
experimentally observed graph ${\cal G}$ with adjacency matrix 
${\cal M}_{\cal G}$, then $p_{ij,A} = M_{{\cal G},ij}$.
 
The average degree $\langle k' \rangle_{k}$ of neighbors of nodes with 
degree $k$ is
\beq
   \langle k' \rangle_{k} = \frac{\sum_{k'} k' N(k,k')}{\sum_{k'} N(k,k')} .
   \label{kkprime}
\eeq 
where we have dropped the subscript ``$A$" for brevity. This 
quantity can be related to the (dis)assortativity, i.e. the tendency of
nodes to connect (less) preferentially to nodes with similar degree. 
The assortativity was formally introduced by Newman as the Pearson
correlation coefficient for the degrees of any two nodes connected
by an edge \cite{newman02}. Intuitively, when the average degree
$\langle k' \rangle_{k}$ is an increasing function of $k$ then the
network shows assortative mixing, i.e. nodes of low degree tend to
connect to nodes of low degree and nodes of high degree tend to
connect to nodes of high degree. When $\langle k' \rangle_{k}$ is flat, 
the network shows no assortativity, and when $\langle k' \rangle_{k}$ 
is a decreasing function of $k$ then the network shows disassortative 
mixing~\cite{catanzaro}.

We can compute $\langle k' \rangle_{k}$ in several PFDS ensembles. The 
ensemble $Z_3(k_1,k_2)$ consists of uncorrelated random graphs with $N$
nodes, $L$ edges and no multiple or self-connections, where we fix
the degrees of one pair of nodes. Evidently, we choose the pair whose link 
likelihood is being evaluated. Eq.~(\ref{kkprime}) is
then calculated by averaging over all pairs of nodes in the network. 
This clearly allows us to take the whole degree sequence into consideration, 
although only pairs of node degrees are considered simultaneously.
To include the presence of a hub, we work in $Z_{4}(k_1,k_2,k_{\rm max})$, the
ensemble of uncorrelated random graphs with $N$ nodes, $L$ edges and
no multiple or self-connections, where we fix the degree of the pair
of nodes whose link likelihood is being evaluated, as well as the
degree $k_{\rm max}$ of the strongest hub.  

\begin{figure}
  \begin{center}
  \psfig{file=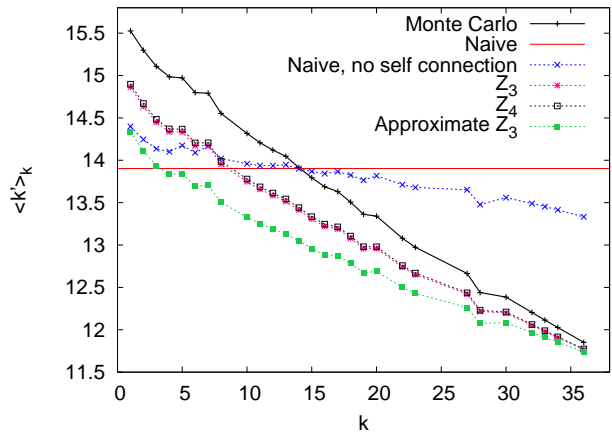,width=6.cm,angle=270}
  \caption{Average degree of the neighbor, 
    $\langle k' \rangle_k$ vs. node degree $k$ for an \textit{Escherichia coli} protein
    interaction network ~\cite{ecoli} in several ensembles.  The FDS  ensemble,
    sampled by Monte Carlo rewiring, shows disassortativity as $\langle k'\rangle_k$ is
    a decreasing function of $k$. For the naive estimate $p_{ij}=k_ik_j/2L$ and using the 
    exact degree sequence, there is no disassortativity (while using a power law as in 
    Eq.~(\ref{pl}) leads to divergence). However, using the
    naive estimate but forbidding self-connection results in slight
    disassortativity.  Approximate $Z_3$, exact $Z_3$, and $Z_{4}(k_1,k_2,k_{\rm max})$ are
    increasingly refined estimates of the FDS ensemble. Notice that the latter two 
    can hardly be distinguished.}
\label{kav}
\end{center}
\end{figure}

As shown in Section V and in Appendix B, we can compute 
$\langle k' \rangle_{k}$ exactly in $Z_3(k_1,k_2)$ and
$Z_{4}(k_1,k_2,k_{\rm max})$, as well as in the approximate $Z_3(k_1,k_2)$ 
ensemble with $p_{ij}$ given by Eq.(\ref{Pkk}) [see also Eq.~(\ref{p2})
below], which becomes exact in the limit of large $N$ for sparse networks,
and for $k_{\rm max}\ll L$. In Fig.~\ref{kav} we plot $\langle k' \rangle_k$ 
versus $k$ for an \textit{Escherichia coli} protein interaction 
network~\cite{ecoli}.  The FDS ensemble, as sampled by the Monte Carlo rewiring
procedure, is clearly disassortative, while the estimate of $\langle k'
\rangle_{k}$ using the standard naive estimate $p_{ij} = k_i k_j/2L$ shows 
no disassortativity or assortativity, as expected. We note that forbidding 
self-connection but otherwise using the naive estimate for $p_{ij}$ results 
in slight disassortativity, while approximate $Z_3$, exact $Z_3(k_1,k_2)$, 
and $Z_{4}(k_1,k_2,k_{\rm max})$ are increasingly refined estimates of the FDS
behavior (see Fig. \ref{kav}).  Finally, $Z_{4}(k_1,k_2,k_{\rm max})$ gives only a slight
improvement over $Z_3$, indicating that hubs \textit{per se} are less important to
global properties such as disassortativity than constraints such as no
self- or multiple connections, which are already implemented at the level of 
$Z_3$, along with information about the whole degree sequence, taken in degree pairs.

The approximate $Z_3$ ensemble is of further interest because $\langle k' 
\rangle_{k}$ can be calculated analytically.  Note that 
\begin{eqnarray}
N(k,k')&=& \sum_{i,j} p_{ij}^{model}\delta(k_i -
  k)\delta(k_j -k') \nonumber \\ 
  &=& P(k,k')N(k)N(k').
\end{eqnarray}
For a power law degree distribution, 
\beq 
    N(k) = (\gamma - 1)Nk^{-\gamma}  
\eeq 
where $N$ is the number of nodes in the network and the degree distribution 
of the network is a power law with exponent $\gamma$.  In this case the sums
over $k'$ in Eq.~(\ref{kkprime}) can be approximated by integrals,
yielding 
\beq 
   \langle k' \rangle_{k} = \frac{\int_{1}^{\infty} k'
   P(k,k')k'^{-\gamma}\;dk' }{\int_{1}^{\infty} P(k,k')
   k'^{-\gamma}\;dk'} \;.      \label{pl}
\eeq 
If we approximate $P(k,k')$ by Eq.~(\ref{Pkk}), these integrals can be 
solved in terms of hypergeometric functions: 
\beq
   \langle k' \rangle_k = \frac{\gamma -1}{\gamma -2} \frac{_2
    F_1(1,\gamma -2; \gamma -1; \frac{-2L}{k})}{_2 F_1(1,\gamma -1;
    \gamma ; \frac{-2L}{k})}\;.
                                       \label{hyper}
\eeq
We note that for $\gamma = 2.5$ the hypergeometric
functions can be expressed in terms of elementary functions:
\beq 
   \langle k' \rangle_k = \frac{\arcsin\left(\sqrt{\frac{2L}{2L + k}}\right)}
   {(\frac{k}{2L})^{1/2} - (\frac{k}{2L})\arctan\left((\frac{k}{2L})^{-1/2}\right)}.
                                    \label{hyper1}
\eeq
To test the validity of Eq.~(\ref{hyper}), we turn to a large network, 
specifically Newman's recent AS-level Internet data \cite{newmanInternet}, 
for which $L = 48436$. In \cite{park03} it is reported that $\gamma 
\approx 2.2 \pm 0.3$ for the Internet; we estimate $\gamma \approx 2.1 
\pm 0.3$ for Ref.~\cite{newmanInternet}.  

\begin{figure}
  \begin{center}
   \psfig{file=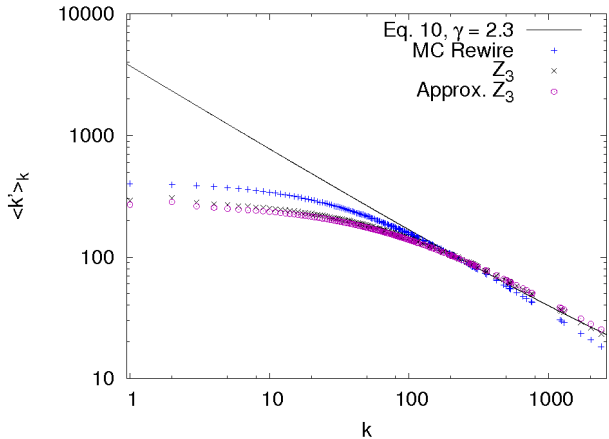,width=6.cm,angle=270}
   \caption{$\langle k' \rangle_k$ versus $k$ for the Internet at the
     AS level from \cite{newmanInternet}.  We plot the analytic estimate of 
     Eq.~(\ref{hyper}) for $\gamma = 2.3$; for comparison we plot the results in
     the approximate $Z_3$ and $Z_3$ ensembles, computed directly from the
     degree sequence of \cite{newmanInternet}, as well as Monte Carlo 
     rewiring estimates of the FDS ensemble.}
   \label{FigInternet}
  \end{center}
\end{figure}

In Fig.~\ref{FigInternet} we plot the analytic estimates of $\langle k'
\rangle_k$ given by Eq.~(\ref{hyper}) for $\gamma = 2.3$. This value for 
$\gamma$ gives the best fit for $\langle k' \rangle_k$ and is within the 
uncertainty of the direct degree distribution measurement of $\gamma$. 
For comparison we also show the results in the approximate $Z_3$ and
$Z_3$ ensembles, computed directly from the degree sequence of
\cite{newmanInternet}, as well as Monte Carlo rewiring estimates of
the FDS ensemble.  Note the strong similarities between the
$Z_3$ results and the Monte Carlo estimates of the FDS ensemble; in 
particular, we observe a flattening of $\langle k' \rangle_k$ for small 
$k$ both in the $Z_3$ ensemble and in the Monte Carlo rewiring. This is 
consistent with the observations of \cite{park03}.  

Also note the similar scaling of the various estimates and of the 
Monte Carlo results at large $k$.  For the Internet, $\langle k' \rangle_k$ 
has been reported to scale with $k$ as a power law, $\langle k' 
\rangle_k \approx k^{-\nu}$ with $\nu\approx 0.5$ \cite{park03}. Our  
Monte Carlo results for the FDS ensemble, using the degree sequence of 
Ref.~\cite{newmanInternet}, show indeed such a power law for large $k$, 
but with $\nu \approx 0.75$. The exact and approximate $Z_3$
calculations, obtained with the exact degree sequence, give $\nu \approx 
0.7$ resp. $0.62$. When using a power law degree sequence and Eq.~(\ref{hyper}), 
the scaling depends on $\gamma$. But in this case, one can verify that 
scaling does not hold in the large $k$ limit, but in the limit $k \ll L$.
The curvature of the continuous line visible in Fig.~\ref{FigInternet}
results entirely from the fact that $k$ is not much less than $L$. 
Thus it is the slope of the continuous line at small $k$ which should 
be used for extracting $\nu$. With this, one finds that $\nu$ varies from 
$\approx 0.79$ for $\gamma = 2.2$ to 0.5 for $\gamma = 2.5$.  Note that $\nu =
0.5$ for $\gamma = 2.5$ is an exact result that can be obtained analytically
by taking the limit $2L/k /to \infty$ in Eq.~(\ref{hyper1}).

In contrast to the approach of
\cite{park03}, all of our results can be computed directly from the
degree distribution or the degree sequence, omitting the intermediate
step of constructing a fugacity distribution to match the statistics
of the degree distribution and then extracting $\langle k' \rangle_k$
from the fugacities.  The fact that the disassortativity properties of
the Internet can be studied so directly in the simple approximate $Z_3$ 
ensemble suggests that Eq.~(\ref{Pkk}) should replace the naive estimate
$P(k,k') = kk'/2L$ in other applications, for example in the study of
motifs.  This is explored in Section V.B (see Eq.~\ref{simple}).

For undirected networks and any null model $A$, Maslov \textit{et al.}~\cite{maslov04} 
defined a quantity called the correlation profile $ R(k,k')=N_{\cal G}(k,k')/
N_A(k,k')$ and the ${\cal Z}$-score ${\cal Z}(k,k') =[N_{\cal G}(k,k')- N_A(k,k')]/
\sigma_A(k,k')$, where $N_A(k,k')$ is defined as in Eq.~(\ref{NA}), 
$N_{\cal G}(k,k')$ is the analogous quantity for the trivial ensemble (with
$p_{ij}$ replaced by $M_{{\cal G},ij}$), and $\sigma^2_A(k,k')$ is the variance
of the number of links connecting nodes with degrees $k$ and $k'$ in 
ensemble $A$ (remember that $N_A(k,k')$ was the average of that number).
The specific null studied in~\cite{maslov04} was the FDS ensemble. As 
shown in Appendix A, a similar analysis can be done comparing different 
null models to each other. Results are also discussed in Appendix A.

\section {Analytic Estimates of the FDS ensemble for Undirected Networks}

\subsection{Notation and Basic Identities}

We now derive our principal analytic results. Our central object 
is the partition function $Z$, which counts the number of graphs 
in the ensemble.  The elementary constraints on the network ($N$ 
nodes, no multiple or self-connections) imply that the adjacency 
matrix $\cal M$ is $N \times N$, is symmetric (for undirected networks), 
has zeroes along the diagonal, and consists solely of $0$'s and $1$'s. 
If we add the constraint of $L$ links, the  partition function can be 
written as
\beq
 Z_1(N,L) = \sum_{\stackrel{\{M_{ij}= 0,1\}}{i<j}} \delta(L - \sum_{i<j}M_{ij})
\eeq
where the sum is over the upper triangle of $\cal M$ due to
symmetry.  A simple computation gives
Eq.(\ref{eq:simple})
for the number of ways to distribute $L$ links among
${N \choose 2}$ possible pairs of nodes.

Now let us specify the degrees of $m$ of the nodes. We refer to $m$ 
as the ``order'' of a calculation. The partition function becomes
\begin{multline}
    Z_{m+1}(N,L,k_{1},...,k_{m}) = \sum_{\stackrel{\{M_{ij}\}}{i<j}} \delta(L -
    \sum_{i<j}M_{ij}) \\
    \times \prod_{l = 1}^{m} \delta(k_{l} -
    \sum_{j=1}^{l-1}M_{j l} - \sum_{j=l+1}^{N}M_{lj})
                 \label{eq:subpartition_undirected_1}
\end{multline}
where we use the symmetry of the adjacency matrix to write the degree 
constraints in terms of the variables $M_{ij}$ with $i <j$.

\begin{figure}
  \begin{center}
  \psfig{file=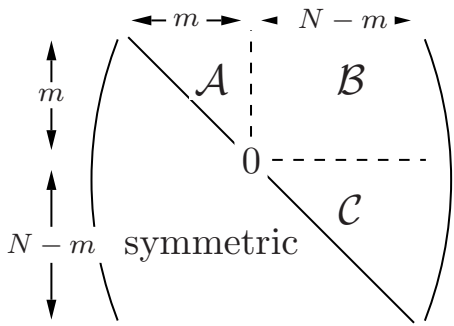,width=6.cm,angle=0}
  \caption{Schematic decomposition of the adjacency matrix $\cal M$
      into the three sub-matrices ${\cal A,B,C}$ discussed in the text. Note
      that $\cal M$ is symmetric for undirected networks.} 
  \label{undirmatrix}
  \end{center}
\end{figure}

It will be helpful at this point  to
introduce some further notation to assist us in organizing this
calculation.  We split the matrix $\cal M$ into four pieces:  
\begin{itemize}
\item ${\cal A}$, the square submatrix controlling the edges linking the 
     $m$ nodes with fixed degree to each other;
\item ${\cal B}$, the rectangular matrix encoding the connections of the $m$ 
     nodes of fixed degree with the rest of the nodes in the  graph, 
     and its transpose ${\cal B}^T$;
\item ${\cal C}$, the square submatrix encoding the edges among the $N-m$
     remaining nodes  (whose degrees are not specified).
\end{itemize}
Due to the symmetry of $\cal M$ only $\cal B$ and the upper triangular 
parts of ${\cal A}$ and ${\cal C}$ are independent. In Fig.~\ref{undirmatrix} 
we present a schematic decomposition of $\cal M$.  

The sum over all adjacency matrices decomposes into a sum over the ${\cal A}$, $\cal B$
and $\cal C$ sub-matrices, with suitable constraints. In particular, we write 
the symbol $\sum_{\{\cal A\}}$ for the sum over all possible values $\{0,1\}$ of
the matrix elements of the submatrix ${\cal A}$. Each term of this sum corresponds 
to a particular (possibly disconnected) labelled subgraph involving $m$ 
nodes of fixed degree.  This is analogous to a diagrammatic expansion of the
partition function, where the partition function is now written as the sum
over all possible subgraphs involving nodes $1$ through
$m$, and each subgraph is weighted by a degeneracy factor resulting 
from the summations over $\cal B$ and $\cal C$. This degeneracy counts the number 
of possible graphs in the ensemble containing that particular subgraph
(respectively submatrix) ${\cal A}$. The partition function is written
in this notation as:
\begin{equation}
    Z_{m+1}= \sum_{\{\cal A\}} Z_{m+1}({\cal A})
\end{equation}
where $Z_{m+1}({\cal A})$ is the partition function or degeneracy factor for a given 
fixed submatrix (equivalently, subgraph) $\cal A$. For each $Z_{m+1}({\cal A})$,
the nodes with fixed degrees (i.e. the first $m$ nodes) are connected in a 
specified way. Thus, for example, the probability of some particular $m\times m$ 
subgraph $\cal A$ occurring would be
\begin{equation}
   {\rm Prob}({\cal A}) = \frac{Z_{m+1}({\cal A})}{Z_{m+1}} \;.
\end{equation}

As shown in Appendix B, the degeneracy of a given subgraph $Z_{m+1}({\cal A})$ 
can be written as
\begin{multline} 
   Z_{m+1}({\cal A}) = {{N-m \choose 2} \choose {L + \sum_{i<j,=2}^{j=m}A_{ij}- \sum_{l=1}^{m}k_{l}}} \\
   \times \prod_{l = 1}^{m}{N-m \choose k_{l} - \sum_{j=1}^{l-1}A_{j l} - \sum_{j=l+1}^{m}A_{lj}}
                      \label{eq:subpartition_undirected}
\end{multline} 

The first term on the right hand side is the degeneracy associated with
the upper half triangle of the square submatrix $\cal C$.  Recall that 
the submatrix $\cal C$ defines connections between all $N-m$ nodes not in 
$\cal A$, i.e. all nodes with free degree. This matrix has ${N-m \choose 2}$ 
independent places to put a specified number of 1's. The number of 1's 
in $\cal C$ depends on $L$, the total number of 1's in the entire (upper 
triangular) adjacency matrix, minus the number of 1's from edges that have 
at least one end on a node of fixed degree. By definition, those
1's cannot appear in $\cal C$. Due to the symmetry of the entire
matrix, the number of 1's to be placed in the upper triangular half of
$\cal C$ is $L + \sum_{i<j,j=2}^{j=m}A_{ij}- \sum_{l=1}^{m}k_{l}$.  

Each factor in the product $\prod_{l = 1}^{m}$ of 
Eq.~(\ref{eq:subpartition_undirected}) is the degeneracy associated with a row
in the matrix $\cal B$. For every such row there are $N-m$ places to put 1's, and
the number of 1's that must be placed in the $l$-th row is the degree of the node
$k_l$ minus the number of 1's in the corresponding row of the entire matrix 
$\cal A$. This latter number is the degree of the node within the subgraph $\cal A$. 

\subsection{Calculation of link likelihoods for undirected networks}

As a first application we compute the link likelihood for two nodes in
this framework.  At lowest order we can just specify the degrees of 
the two nodes under consideration, giving the ensemble $Z_3$ with $m=2$. 
For convenience, the two nodes are labelled $1$ and $2$. The two possible 
configurations of this subsystem, where the nodes are either connected 
or disconnected, must be weighted by the appropriate degeneracy factors according
to Eq.~\ref{eq:subpartition_undirected}. Let us denote these two configurations 
by $\aleph$ and $\aleph'$. In $\aleph$ an edge connects the two nodes, 
so $\aleph_{12}= 1$.  We denote this ``connected'' part of the total partition 
function $Z_{3}^{conn} = Z_3(\aleph)$. In $\aleph'$ there is no edge 
between the two nodes, so $\aleph'_{12}=0$. We denote this ``disconnected'' 
part of the total partition function $Z_{3}^{disc} = Z_3(\aleph')$.  
The total partition function is thus $Z_{3}^{total} = Z_{3}^{conn} + 
Z_{3}^{disc} = Z_{3}(\aleph) + Z_{3}(\aleph')$, thus
\beq
   p^{(m=2)}_{12} = \frac{Z_{3}^{conn}}{Z_{3}^{total}} = 
      \frac{Z_3(\aleph)}{Z_3(\aleph) + Z_3(\aleph')}\;.
\eeq
From Eq.~\ref{eq:subpartition_undirected}, the explicit expressions are
\beq
   Z_3(\aleph) = {{N-2 \choose 2} \choose L + 1 - k_1 - k_2 }
        \prod_{l = 1}^{2}{N-2 \choose k_{l} - 1} \quad ,
\eeq
\beq
   Z_3(\aleph') = {{N-2 \choose 2} \choose L - k_1 - k_2 }
         \prod_{l = 1}^{2}{N-2 \choose k_{l} } \quad .
\eeq
Straightforward calculation gives 

\begin{eqnarray}
   p&&\!\!\!\!\!\!\!^{(m=2)}_{12}    \\
   &&\!\!\!\!\!\!= \left[ 1+\frac{(L+1-k_1-k_2)(N-1-k_1)(N-1-k_2)}{k_1k_2((N-2)(N-3)/2-L+k_1+k_2)}\right]^{-1}
         \nonumber            \label{pbig}
\end{eqnarray}
and in the limit $L\to \infty$, $L/N^2 \to 0$, and $k_i \ll L$ this reduces to 
\beq
    p^{(m=2)}_{12} \approx \frac{k_1 k_2}{2L + k_1 k_2}. 
             \label{p2}
\eeq
In the limit $k_1 k_2 << L$ we recover from Eq.~(\ref{p2}) the naive estimate $k_1 k_2/2L$  
used by most authors. This naive estimate is a bad approximation if either node 1 or 
node 2 is a hub. In general, the full expression given in Eq.~(21) is a better 
approximation, although as we have shown in section IV the approximate $Z_3$ ensemble 
given by Eq.~(\ref{p2}) is both analytically tractable and significantly better than 
the naive estimate.

\begin{figure}
  \begin{center}
   \psfig{file=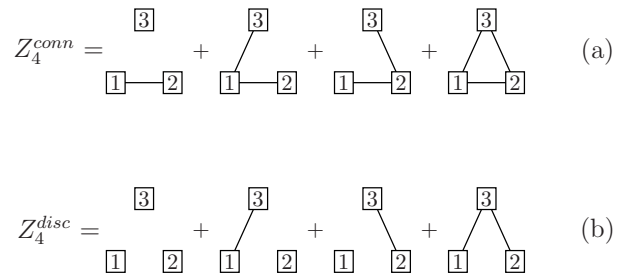,height = 3.6cm,width=8.cm,angle=0}
   \caption{Diagrammatic expansion for the connected (a) and disconnected (b)
      contributions to the partition function $Z_4$ for $p_{12}$.} 
\label{diagram_1}
\end{center}
\end{figure}

The presence of any hubs in the network reduces the link likelihood between two nodes, 
particularly nodes of low degree, as
their links are ``stolen" by the hubs. This effect already appears in the calculation 
of the disassortativity, as shown in Figures \ref{kav} and \ref{FigInternet}. We can 
refine the preceding computation
by including constraints on the degree sequence coming from the hubs. We incorporate
these constraints from the hubs by considering the nodes with the highest degrees first.
The partially fixed degree sequence ensemble $Z_4(k_1,k_2,k_{\rm max})$, then, 
includes the two nodes $k_1,k_2$ whose link likelihood we compute (free to vary over 
all node pairs) and the largest hub in the network, with degree ($k_{\rm max}$). In 
the case that $k_{\rm max} = k_1$ or $k_2$ we use the next largest hub degree for the 
third constraint. In the sub-matrices $\cal A$, all possible ways to connect the three 
nodes are enumerated.  To compute the link likelihood $p^{(m=3)}_{ij}$ we divide the 
partition function into connected $Z_4^{conn}$ and
disconnected $Z_4^{disc}$ parts, where connected means an edge connects 
node 1 and node 2, i.e. $A_{12}=1$.  The diagrammatic expansion for the connected 
sub-partition function is shown in Fig.~\ref{diagram_1}a. and the disconnected 
sub-partition function is shown in Fig.~\ref{diagram_1}b.

\begin{figure}
  \begin{center}
   \psfig{file=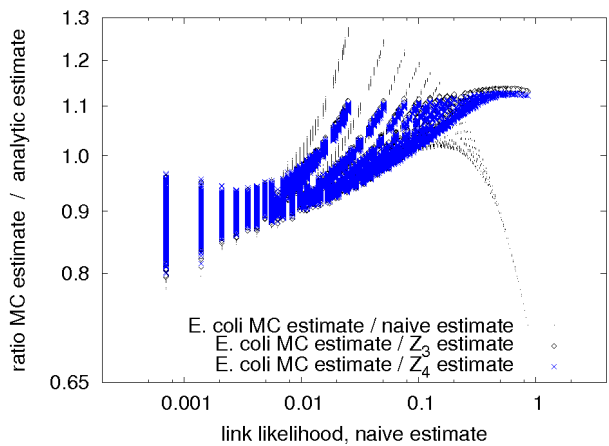,width=6.cm,angle=270}
   \caption{Scatter plot comparing estimates of the link likelihood in an 
     \textit{Escherichia coli} protein interaction network. On the horizontal
     logarithmic axis we plot the naive estimate $\frac{k_{i}k_{j}}{2L}$; on
     the vertical logarithmic axis we plot for all corresponding nodes the 
     ratio of the Monte-Carlo rewiring estimate to the naive, $Z_{3}$ (no hub),
     and $Z_{4}$ (one hub) estimates.  The latter two nearly coincide.} 
\label{undircompare}
\end{center}
\end{figure}

So we can compute the link likelihood as
\begin{eqnarray}
   p^{(m=3)}_{12} &=& \frac{Z_4^{conn}}{Z_4^{conn}+Z_4^{disc}} \\
                  &=& \frac{Z_4(\aleph_1) + Z_4(\aleph_2) + Z_4(\aleph_3) + 
                      Z_4(\aleph_4)}{\sum_{i=1}^8 Z_4(\aleph_i)} \quad ,  \nonumber
\end{eqnarray}
where $\aleph_1$ to $\aleph_8$ denote the adjacency matrices for the eight graphs
shown in Fig.~\ref{diagram_1}.
In Fig.~\ref{undircompare} we compare the link likelihood $p_{ij}$ for an undirected
network in the FDS ensemble (obtained numerically by the rewiring method) to 
the analytic results in the ensembles $Z_3(k_1,k_2)$ and $Z_4(k_1,k_2,k_{\rm max})$. 
For each pair $(i,j)$ we plot the naive estimate 
$\frac{k_{i}k_{j}}{2L}$ on the horizontal axis; the vertical coordinate is the 
ratio of the numerical (Monte-Carlo) estimate to the naive, no hub, and one hub 
estimates for that pair. The $Z_{3}$ estimate is already a substantial improvement 
over the naive analytic estimate, with slight further refinement coming as 
expected from the inclusion of hubs in all the diagrams.

\begin{table*}
\begin{center}
\caption{Comparing estimates of the total number of triangles in various 
networks with $N$ nodes.  MC refers to Monte Carlo rewiring estimates of 
the FDS ensemble.  As expected, the results of Eq.~\ref{complex} approach 
the asymptotic result, Eq.~(\ref{simple}), for large, sparse networks.}
\begin{tabular}{|c| |c| |c|c|c|c|c|c|}
	\hline
\rm{Network} & $N$ &Eq.~(\ref{simple}) & Eq.~(\ref{complex}) & MC & \% Error &  \% Error & \% Error \\
 & & & & & Eq.~(\ref{simple}) vs. Eq.~(\ref{complex}) & Eq.~(\ref{simple}) vs. MC & Eq.~(\ref{complex}) vs. MC \\
	\hline \hline 
\textit{E. coli} &    230 &    215.82 &    289.65 &    322.14 &  25.49 &  33.01 &  10.09 \\
Yeast(narrow)    &   1373 &    302.77 &    247.65 &    339.10 & -22.26 &  10.71 &  26.97 \\
Yeast            &   1373 &    651.07 &    592.59 &   1160.39 &  -9.87 &  43.89 &  48.93 \\
Yeast (broad)    &   1373 &   1553.94 &   1667.70 &   2813.37 &   6.82 &  44.77 &  40.72 \\
AS Internet      &  22963 &  29157.38 &  31840.23 &  37810.68 &   8.43 &  22.89 &  15.79 \\
WWW              & 325729 & 379371.15 & 379706.63 & 274926.89 &   0.09 & -37.99 & -38.11 \\
\hline
\end{tabular}
\end{center} 
\end{table*}

\subsection{Calculation of subgraph likelihoods for undirected networks}

Estimating the likelihoods of larger subgraphs can be done along exactly the 
same lines.  As an example, we estimate the number of triangles in an undirected 
network and test this estimate on an \textit{Escherichia coli} protein 
interaction network \cite{ecoli}, a yeast (\textit{Saccharomyces cerevisiae}) 
protein interaction network \cite{yeast}, two artificial yeast protein 
interaction networks created by modifying the degree sequence of \cite{yeast} 
by hand to make it narrower or broader, the Newman AS level map of the 
Internet~\cite{newmanInternet}, and a symmetrized snapshot of the World Wide 
Web \cite{WWW}.  Working in an ensemble with four constraints, $Z_4(k_1,k_2,k_3)$, 
we consider all permitted triples of nodes $(i,j,k)$, forbidding self- and 
multiple connection.  Note that node 3 is no longer fixed as the largest hub, 
but allowed to range over all nodes.  Given fixed nodes 1, 2, and 3,  we can 
compute the likelihood of a triangle between them as 
\begin{eqnarray}
   p^{(m=3)}_{\Delta} = \frac{Z_4(\aleph_4)}{\sum_{i=1}^8 Z_4(\aleph_i)} 
              \label{complex}
\end{eqnarray}
where $\aleph_4$ corresponds to the fully connected subgraph, i.e. the last 
term in the sum of Fig.~\ref{diagram_1}(a).  The resulting combinatorial 
expression is quite unwieldy.  However, in the limit $L\to \infty$, $L/N^2 \to 0$, 
and $1 \ll k_i \ll L$, i.e. the large, sparse network limit used in deriving 
Eq.~(\ref{p2}) with the additional assumption of $1 \ll k_i$, we find a 
remarkable simplification.  The expression factorizes to
\begin{equation}
   p^{(m=3)}_{\Delta} \approx P(k_1,k_2)P(k_1,k_3)P(k_2,k_3)
\label{simple}
\end{equation}
where $P(k_1,k_2)$ is given by Eq.~(\ref{Pkk}).

We now test these formulae against the various trial networks.  The results are 
shown in Table I, where ``MC" represents the averaged triangle count for many 
Monte Carlo rewirings, i.e. a numerical estimate of the average number of 
triangles in the FDS ensemble.  The only noticeable trend is the decrease in 
the absolute value of the \% Error (defined as [Eq.~(\ref{complex}) - 
Eq.~(\ref{simple})]/Eq.~(\ref{complex})) between the simple factorized expression 
of Eq.~(\ref{simple}) and the elaborate expression of Eq.~(\ref{complex}) as $N$ 
increases.  This verifies the approximation made in deriving Eq.~(\ref{simple}).  
The time required to make the ``complex" estimate given by Eq.~\ref{complex} is 
roughly equal to the time required to count the number of triangles for a single 
Monte Carlo instance; the simple estimate given by Eq.~(\ref{simple}) is much 
faster, running on the WWW data in roughly $90$ seconds (on an Intel core duo processor) 
without any optimization for speed. 

It is tedious to verify similar factorizations 
for larger motifs. However, the convergence to Eq.~(\ref{simple}) in the predicted 
limit provides further evidence for the replacement of the naive estimate 
$P(k,k') = kk'/2L$ used e.g. in \cite{itzkovitz03}, where factorizations as 
in Eq.~(\ref{simple}) were assumed, with Eq.~(\ref{Pkk}). This will be the topic 
of future work.  Such extensions of our method can also be used to study larger 
motifs in complex networks \cite{mayakim, mayapeterkim}, or to study large 
networks where computational time for rewiring grows prohibitively,
but the approximation underlying Eqs.~(\ref{p2}) and (\ref{simple}) should 
still be valid.


\section{Conclusion}
Detecting and describing local structure is an important frontier in the 
study of complex networks, as many of the features distinguishing real-world 
networks from their random analogues or null models are local: degree-degree 
correlations, motifs, and so forth.  One of the major obstacles to 
this project is the lack of analytical techniques to study the fixed 
degree sequence ensemble, which is the most common null model for 
complex networks associated with the rewiring method.  In this paper we 
have reviewed the numerical tools for studying the FDS ensemble 
and discussed some of the practical uses (e.g. disassortativity, motif 
calculation, correlation profiles) to which knowledge of local structure 
can be put.  Through a careful study of the partition function of the FDS 
ensemble and the PFDS ensembles containing it, we derive simple and 
general combinatorial expressions that improve naive estimates of the
link likelihood by explicitly including important constraints from the FDS 
ensemble (the exclusion of multiple edges and self connections, and the 
appearance of a broad range of degrees) in a ``Gaussian" type approximation 
where the set of degree constraints are treated minimally but non-trivially. 

In particular, for undirected networks we have developed the analytically 
tractable approximate $Z_3$ ensemble, where the link likelihood $P(k,k') = 
kk'/(2L + kk')$ (Eq.~(\ref{Pkk})) gives clear disassortativity, while the 
naive estimate $kk'/2L$ does not.  We have also introduced a diagrammatic 
expansion of the PFDS partition function, which organizes the combinatorial 
calculations usefully and leads to a simple, approximate factorized formula 
for the estimated number of undirected triangles between three nodes, 
$P_{\Delta}^{(m=3)}(k_1,k_2,k_3) = P(k_1,k_2)P(k_1,k_3)P(k_2,k_3)$ 
(Eq.~(\ref{simple})) where $k_1,k_2,k_3$ are the degrees of the $3$ nodes. 
The factorization suggests the application of Eq.~(\ref{p2}) to extended 
local structures such as motifs. 

It should be emphasized that these 
analytic results are not merely useful for the null model they have been 
explicitly developed to approximate (the FDS ensemble). They also provide 
guidance in developing more complicated null models that incorporate 
higher-level constraints.  The astronomical $\cal{Z}$-scores observed in 
work on extended motifs \cite{mayakim,mayapeterkim} dramatize the need 
for such extensions, which might constrain, for example, the number of 
triangles in addition to the number of nodes, number of links, and a few 
degrees.  Further work will explore applications of the approximate $Z_3$ 
and other PFDS ensembles to motif estimates, as well as the incorporation 
of higher-level constraints in the PFDS ensemble to improve likelihood 
estimates of extended motifs.  The extension of the results of this paper 
to directed networks is in preparation.      

\section*{Acknowledgements}
J.G.F thanks Amer Shreim for creating the correlation profile in 
Figure~\ref{correlation_3}. We thank Gabe Musso for providing the data set 
of Ref.~\cite{yeast}. Part of this work was completed at the 
Perimeter Institute, for which hospitality the authors express their 
sincere thanks.  J.G.F. gratefully acknowledges the support of the 
Rhodes Trust and the Alberta Ingenuity Fund. P.G. thanks iCORE for 
financial support.

\section{Appendix A:  Comparing null models via the correlation profile}

We can study how the real network deviates from various null hypotheses 
by calculating $R(k,k')$ with respect to various null hypotheses. This 
provides an overall measure of how close the ensembles are to each other 
and helps establish the relevant features that distinguish the real
network from the different ensembles.

\begin{figure}
  \begin{center}
   \psfig{file=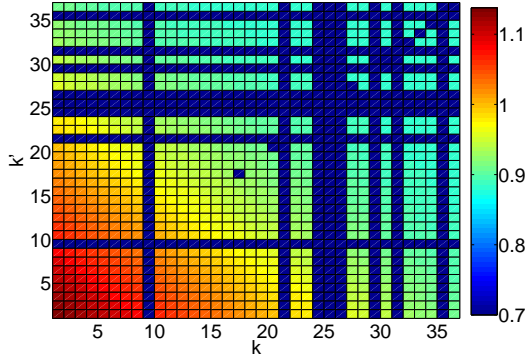,width=7.cm,angle=0}
   \caption{Correlation profile $R_{Z_3|\rm{FDS}}(k,k')$ for the 
     \textit{Escherichia coli} protein interaction network \cite{ecoli}.
     Note that the dark blue regions are artificially set to the value 
     $.7$; they correspond to values of $(k,k')$ for which no data exist.} 
\label{correlation_3}
\end{center}
\end{figure}

In general, we can define correlation profiles and ${\cal Z}$-scores for 
any pair $\left(A,B \right)$ of ensembles:
\beq
    R_{A|B}(k,k') = \frac{ N_{A}(k,k')}{N_{B}(k,k')}.  
\eeq 
and
\beq 
    {\cal Z}_{A|B}(k,k') = \frac{N_{A}(k,k') - N_{B}(k,k')}{\sqrt{\sigma_A^2 + \sigma_B^2}}.
\eeq 
In particular, we can calculate the correlation profile for the numerically 
sampled FDS ensemble and the ensemble $Z_3(k_1,k_2)$.  Recall that
the ensemble $Z_3$ consists of uncorrelated random graphs with $N$
nodes, $L$ edges and no multiple or self- connections, where we fix
the degree of the pair of nodes whose link likelihood is being evaluated. 
We might also compare the FDS ensemble to $Z_{4}(k_1,k_2,k_{\rm max})$, 
the ensemble of uncorrelated random graphs with $N$ nodes, $L$ edges and
no multiple or self-connections, where we fix the degree of the pair
of nodes whose link likelihood is being evaluated, as well as the
degree of the largest node in the network, $k_{\rm max}$.  In $Z_3$ and
$Z_4$, $p_{ij}$ can be calculated exactly (see Section V). We plot
$R_{Z_3|\rm{FDS}}(k,k')$ for the \textit{Escherichia coli} protein 
interaction network in Fig.~\ref{correlation_3}.  It is clear that the 
$Z_3$ ensemble captures many of the features of the FDS ensemble, as
$R_{Z_3|\rm{FDS}}(k,k')$ is close to $1$ for all $k,k'$.
$R_{Z_4|\rm{FDS}}(k,k')$ exhibits slight improvement over $Z_3$, as
expected (data not shown).  The correlation profile allows us to 
identify correlations due to the degrees of the other nodes in the 
network and provides a test of our hypothesis that the PFDS
ensemble captures much of the structure of the FDS ensemble.

\begin{widetext}
\section{Appendix B: Derivation of $Z_{m+1}$ for Undirected networks}

We note that the number of possible undirected subgraphs of $m$ nodes is $2^{(m^{2} - m)/2}$.  
So we write:

\begin{align}
Z_{m+1}(N,L,k_{1},...,k_{m}) &=  \nonumber \\
&\sum_{\{A_{ij}\}} \sum_{\{B_{ij}\}}
    \sum_{\{C_{ij}\}} \delta(L -(\sum_{\stackrel{i<j}{j=2}}^{j=m}A_{ij})-(\sum_{\stackrel{i \leq m}{j=m+1}}^{j=N}B_{ij})-(\sum_{\stackrel{i<j}{i=m+1}}^{i=N}C_{ij}))
\nonumber \\
&\times \prod_{l = 1}^{m} \delta(k_{l} -
    \sum_{j=1}^{l-1}A_{j l} - \sum_{j=l+1}^{m}A_{lj} - \sum_{j=m+1}^{N}B_{lj})
\end{align}

For concision we henceforth write the sum more compactly, as in the next equation.

We now Fourier transform the delta-functions:
\begin{align}
Z_{m+1}(N,L,k_{1},...,k_{m}) &= \sum_{\{\cal ABC\}} \frac{1}{(2\pi)^{m+1}}\int_{-\pi}^{\pi}...\int_{-\pi}^{\pi} dz da_{1}...da_{m} e^{\imath z(L -\sum_{\stackrel{i<j}
    {j=2}}^{j=m}A_{ij})}
\nonumber \\ 
&\times \prod_{l = 1}^{m} e^{\imath a_{l}(k_{l} -
    \sum_{j=1}^{l-1}A_{j l} - \sum_{j=l+1}^{m}A_{lj})} \prod_{l=1}^{m}\prod_{j=m+1}^{N}
    e^{-\imath(z+a_{l})B_{lj}}\prod_{\stackrel{i<j}{i=m+1}}^{N}e^{-\imath z
    C_{ij}}
\end{align}

We can do the sum over $B_{ij}$, yielding:
\begin{align}
Z_{m+1}(N,L,k_{1},...,k_{m})&= \sum_{\{\cal AC\}}
 \frac{1}{(2\pi)^{m+1}}\int_{-\pi}^{\pi}...\int_{-\pi}^{\pi} dz da_{1}...da_{m} e^{\imath z(L -\sum_{\stackrel{i<j}
    {j=2}}^{j=m}A_{ij})} \nonumber \\
&\times \prod_{l = 1}^{m} e^{\imath a_{l}(k_{l} -
    \sum_{j=1}^{l-1}A_{j l} - \sum_{j=l+1}^{m}A_{lj})}
    \prod_{l=1}^{m}(1+ e^{-\imath(z+a_{l})})^{N-m}\prod_{\stackrel{i<j}{i=m+1}}^{N}e^{-\imath z
    C_{ij}}
\end{align}

Performing the standard binomial expansion yields:
\begin{align}
Z_{m+1}(N,L,k_{1},...,k_{m}) &= \sum_{\{\cal AC\}} \frac{1}{(2\pi)^{m+1}}\int_{-\pi}^{\pi}...\int_{-\pi}^{\pi} dz da_{1}...da_{m} e^{\imath z(L -\sum_{\stackrel{i<j}
    {j=2}}^{j=m}A_{ij})} \nonumber \\
&\times \prod_{l = 1}^{m} e^{\imath a_{l}(k_{l} -
    \sum_{j=1}^{l-1}A_{j l} - \sum_{j=l+1}^{m}A_{lj})}
    \prod_{l=1}^{m}\sum_{n_{l}=0}^{N-m}{N-m \choose n_{l}} e^{-\imath n_{l}(z+a_{l})}\prod_{\stackrel{i<j}{i=m+1}}^{N}e^{-\imath z
    C_{ij}}
\end{align}

Integrating over $a_{1}...a_{m}$ sets $n_{l} = k_{l} -
    \sum_{j=1}^{l-1}A_{j l} - \sum_{j=l+1}^{m}A_{lj}$, so we have:
\begin{align}
Z_{m+1}(N,L,k_{1},...,k_{m}) &= \sum_{\{\cal AC\}}\frac{1}{(2\pi)}\int_{-\pi}^{\pi}dz e^{\imath z(L -\sum_{\stackrel{i<j}
    {j=2}}^{j=m}A_{ij})}\nonumber \\
    &\times \prod_{l = 1}^{m} e^{-\imath z(k_{l} -
    \sum_{j=1}^{l-1}A_{j l} - \sum_{j=l+1}^{m}A_{lj})}{N-m \choose k_{l} -
    \sum_{j=1}^{l-1}A_{j l} - \sum_{j=l+1}^{m}A_{lj}}
    \prod_{\stackrel{i<j}{i=m+1}}^{N}e^{-\imath z C_{ij}}
\end{align}

We now sum over ${C_{ij}}$ and perform the binomial expansion of the
resulting quantity:
\begin{align}
Z_{m+1}(N,L,k_{1},...,k_{m}) & = \sum_{\{\cal A\}}
    \frac{1}{(2\pi)}\int_{-\pi}^{\pi}dz e^{\imath z(L + \sum_{\stackrel{i<j}
    {j=2}}^{j=m}A_{ij}- \sum_{l=1}^{m}k_{l})}\nonumber \\
    &\times \prod_{l = 1}^{m}{N-m \choose k_{l} -
    \sum_{j=1}^{l-1}A_{j l} - \sum_{j=l+1}^{m}A_{lj}}
    \sum_{n=0}^{{N-m \choose 2}}{{N-m \choose 2} \choose n}e^{-\imath zn}
\end{align}
where we have used the fact that
$$\prod_{l = 1}^{m} e^{\imath z(\sum_{j=1}^{l-1}A_{j l} +
   \sum_{j=l+1}^{m}A_{lj})} = e^{\imath
   z(\sum_{l=1}^{m}(\sum_{j=1}^{l-1}A_{j l} +
   \sum_{j=l+1}^{m}A_{lj}))}= e^{2\imath z \sum_{\stackrel{l<j}
    {j=2}}^{j=m}A_{lj}}$$
by the symmetry of the adjacency matrix.  We may now integrate over
$z$ to give the result:
\begin{equation}
    Z_{m+1}(N,L,k_{1},...,k_{m}) = \sum_{\{{\cal A}\}}
    {{N-m \choose 2} \choose (L + \sum_{i<j,j=2}^{j=m}
    A_{ij}- \sum_{l=1}^{m}k_{l})} \prod_{l = 1}^{m}{N-m \choose k_{l} -
    \sum_{j=1}^{l-1}A_{j l} - \sum_{j=l+1}^{m}A_{lj}}
\end{equation} 
This can be written as a sum over sub-partition sums $Z_{m+1}({\cal A})$, 
each of which is given by Eq.~(\ref{eq:subpartition_undirected}). Thus 
we recover the result for fixed $\cal A$ given in Section V.

\end{widetext}

%


\begin{thebibliography}{30}
\bibitem{footnote1} This is the number of ways of placing $L$ identical objects
  (in this case ``1"s) in $N(N-1)/2$ places.
\bibitem{footnote2} In many cases one wants to construct a surrogate ensemble with the same
  degree sequence as a given graph $\cal G$. If a given graph $\cal G$ is available, one does not
  need this first step, and starts from $\cal G$ directly.
\end{thebibliography}
\end{document}